# Sub-0.6 eV Inverted Metamorphic GaInAs Cells Grown on InP and GaAs Substrates for Thermophotovoltaics and Laser Power Conversion

Kevin L. Schulte,[1,a)] Daniel J. Friedman,[1] Titilope Dada,[2] Harvey L. Guthrey,[1] Edgard Winter da Costa,[1] Eric J. Tervo,[2] Ryan M. France,[1] John F. Geisz,[1] Myles A. Steiner[1]

[1]*National Renewable Energy Laboratory, Golden, Colorado 80401, USA*

[2]*University of Wisconsin-Madison, Department of Electrical and Computer Engineering, Madison, Wisconsin 53706, USA*

[a)]**Author to whom correspondence should be addressed**: kevin.schulte@nrel.gov

**Abstract**

We present inverted metamorphic $Ga_{0.3}In_{0.7}As$ photovoltaic converters with sub-0.60 eV bandgaps grown on InP and GaAs substrates. The compositionally graded buffers in these devices have threading dislocation densities of $1.3\pm0.6 \times 10^6$ cm$^{-2}$ and $8.9\pm1.7 \times 10^6$ cm$^{-2}$ on InP and GaAs, respectively. The devices generate open-circuit voltages of 0.386 V and 0.383 V, respectively, at a current density of ~10 A/cm$^2$, yielding bandgap-voltage offsets of 0.20 and 0.21 V. We measured their broadband reflectance and used it to estimate thermophotovoltaic efficiency. The InP-based cell is estimated to yield 1.09 W/cm$^2$ at 1100 °C vs. 0.92 W/cm$^2$ for the GaAs-based cell, with efficiencies of 16.8 vs. 9.2%. The efficiencies of both devices are limited by sub-bandgap absorption, with power weighted sub-bandgap reflectances of 81% and 58%, respectively, which we assess largely occurs in the graded buffers. We estimate that the thermophotovoltaic efficiencies would peak at ~ 1100 °C at 24.0% and 20.7% in structures with the graded buffer removed, if previously demonstrated reflectance is achieved. These devices also have application to laser power conversion in the 2.0-2.3 µm atmospheric window. We estimate peak LPC efficiencies of 36.8% and 32.5% under 2.0 µm irradiances of 1.86 W/cm$^2$ and 2.81 W/cm$^2$, respectively.

## 1. Introduction

Emerging applications such as thermal energy storage,[1] waste heat recovery,[2],[3] small modular nuclear reactors,[4] and portable power generation[5] require thermophotovoltaic (TPV) converters that collect wavelengths longer than 2 µm. Some thermal energy storage applications using TPVs operate at high temperatures (>1500 °C) and can use devices with designs adapted from photovoltaics applications,[6] but there are also potentially impactful applications, for example waste heat recovery, for converters with significantly lower bandgaps to access heat sources at lower temperatures around 1000 °C or lower. Interest in laser power beaming is also growing, and there is an atmospheric window from 2.0-2.3 µm[7] where high efficiency laser power converters (LPCs) operating in that bandgap range could be useful when coupled with Ho:YAG or Tm fiber lasers emitting at these wavelengths.[8] Converters using the quaternary alloy $Ga_xIn_{1-x}As_ySb_{1-y}$, lattice-matched to a GaSb substrate are one option at these wavelengths, but challenges in the epitaxy of these materials limit their performance.[9-10] Metamorphic $Ga_xIn_{1-x}As$ alloy enables access to bandgaps between 1.42 and 0.35 eV and is a common choice for these applications because it can be grown epitaxially with high quality.[9] 0.74 eV $Ga_{0.47}In_{0.53}As$ devices are grown lattice-matched to InP,[11] while lower bandgaps require graded buffers to shift the lattice constant and access metamorphic GaInAs. Upright photovoltaic cells with bandgaps as low as ~0.5 eV were developed on InP using $InP_yAs_{1-y}$ or $Ga_xIn_{1-x}As$ compositionally graded buffers.[12-15]

A crucial component of achieving high TPV efficiency is to return unused, sub-bandgap photons, which do not contribute to PV power generation, back to the thermal source. Instead of passing through the device unabsorbed, or being absorbed as a parasitic loss in the device, these photons are reflected back to the emitter, reducing transmission loss and reheating the emitter. An elegant approach to doing this, which has been long understood and practiced,[15] is to implement broadband reflectors behind thin-film devices. Recently, major increases in TPV efficiency were achieved by combining this approach with methods for achieving very high above-bandgap photovoltaic performance.[16-18]. Previous <0.74 eV metamorphic $Ga_xIn_{1-x}As$ cells used upright device structures grown on InP and were left on substrate, with a reflector placed behind the substrate.[15],[19] Inverted growth enables direct access to the rear surface of the device, which permits the application of high reflectivity metals or dielectrics while removing the necessity to use semi-insulating substrates with the attendant processing complexities.[15] Inverted growth is also compatible with epitaxial liftoff techniques, which enable removal of the device from the substrate. The device can be placed on a flexible handle and/or heat sink, and the substrate can potentially be reused to save cost. Furthermore, the graded buffer can be removed[17] to reduce sub-bandgap absorption, which is another critical parameter affecting TPV efficiency.[11] Alternatively, the graded buffer could be left in the device to improve current spreading in high-power LPCs.

Growth of these devices on a substrate other than InP, such as GaAs, is desirable for the following reasons. InP is generally available in diameters up to 100 mm, whereas GaAs is available in sizes up to 200 mm, making GaAs more scalable at the high volumes needed for applications like thermal grid storage. GaAs costs significantly less than InP on an areal basis, and we estimate that a 0.6 eV GaInAs cell grown on GaAs would be 4-5x less expensive than one grown on InP even considering the added cost of the additional metamorphic grading.[20] One more advantage is that GaAs is considerably less brittle than InP, which potentially reduces yield losses due to breakage. Lastly, we note that GaAs has nearly the same lattice constant as Ge substrates, meaning that any device recipes developed on GaAs would be generally compatible with Ge substrates, which are available up to 300 mm diameter.[21] These factors are significant drivers of manufacturability and cost reduction, if these devices can be grown on GaAs with sufficiently high performance.

In this work, we study the inverted metamorphic growth of 0.58-0.59 eV $Ga_{0.30}In_{0.70}As$ photovoltaic cells on InP and GaAs substrates employing broadband rear reflectors. We evaluate the effect of the different lattice mismatch, 1.2% on InP substrates vs. 5.0% on GaAs substrates, on the defect structure of the graded buffers used to bridge the lattice constant gap between the substrate and device layers. We compare the performance of these devices under high irradiance and use these high-intensity data to estimate their TPV efficiency, $\eta_{TPV,}$ as a function of black body emitter temperature and discuss a path to $\eta_{TPV}$ > 24% at 1100 °C. Lastly, we estimate their LPC efficiency, $\eta_{LPC}$, using an idealized 2.0 µm emitter.

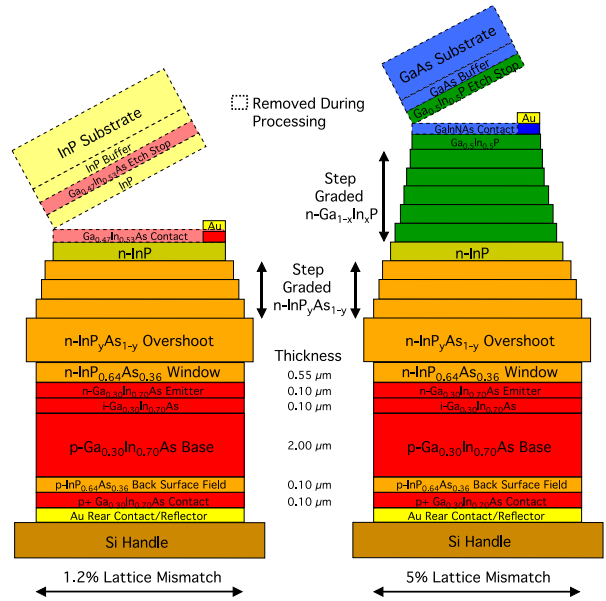

Figure 1. Structure of devices grown on InP and GaAs. The width of the layers indicates the relative lattice constant, i.e. wider layers have a longer lattice constant, and layers with different widths are lattice-mismatched. Color roughly indicates relative bandgap. There are more steps in the actual graded buffers than depicted here. The growth direction is from top to bottom.

## 2. Results and Discussion

### 2.1 Structural Characterization

The device structures for the ~0.58-0.59 eV $Ga_{0.30}In_{0.70}As$ devices grown on InP and GaAs substrates are shown in Figure 1. The $Ga_{0.30}In_{0.70}As$ pn junction where the carrier collection occurs is identical in both devices, but the device on GaAs utilizes a thicker graded buffer to bridge the significantly larger lattice mismatch (5.0%, vs. 1.2% on InP). The first portion of that graded buffer utilizes the $Ga_xIn_{1-x}P$ alloy from $Ga_{0.5}In_{0.5}P$ to InP, and then the buffer switches to an $InP_yAs_{1-y}$ alloy for the remainder of the grading. The InP-based device only uses the $InP_yAs_{1-y}$ portion of the graded buffer. See the Experimental Section for complete growth details.

First, we characterized the surface defect structure of the two types of graded buffers. Figure 2 shows atomic force microscopy (AFM) images, plan view scanning electron microscopy (SEM) images and panchromatic cathodoluminescence (CL) images for InP-based [(a)-(c)] and GaAs-based [(d)-(f)] grades. Both grades exhibit cross-hatched surfaces, characteristic of mismatched epitaxy employing compositionally graded buffers. The GaAs based grade is significantly rougher, with a root mean square (RMS) roughness of 44.6 nm in a 70 µm x 70 µm square scan vs. 18.5 nm for the graded buffer grown on InP. We note that the near-normal reflectance of the surface measured by an *in situ* optical tool decreased significantly during the growth of the $InP_yAs_{1-y}$ portion of that sample, implying that most of the roughening occurred during the growth of those layers. The InP-based grade exhibits a threading dislocation density (TDD) of $1.3\pm0.6 \times 10^6$ cm$^{-2}$ which compares quite favorably to TDD reported in similar graded buffers.[22] The TDD is almost an order of magnitude higher in the GaAs based grade, with TDD = $8.9\pm1.7 \times 10^6$ cm$^{-2}$. This is ~2-3x higher than we observe in GaInP grades from GaAs to InP,[23] suggesting that TDD increase occurred during growth of the InP$_y$As$_{1-y}$ portion of the grade between InP and final lattice constant. Surface roughness correlates with higher TDD,[24-25] either by actively creating drag on gliding dislocations, reducing the amount of strain relaxation per TD, or as a sign of the presence of compositional variation (phase separation) which also restricts glide.[26-27] Therefore, given that phase separation is not expected[28] or observed[22] in InP$_y$As$_{1-y}$, it is likely that the increased roughening that occurred during the grade of the GaAs-based sample reduced the velocity of gliding dislocations, increasing their density. Comparing the SEM and CL images in panels 2e and 2f, it appears that the TD spots are concentrated in the troughs in the rough surface, suggesting that they have been trapped there due to the roughness. Managing this roughening behavior will be key to reducing TDD in these grades in the future.

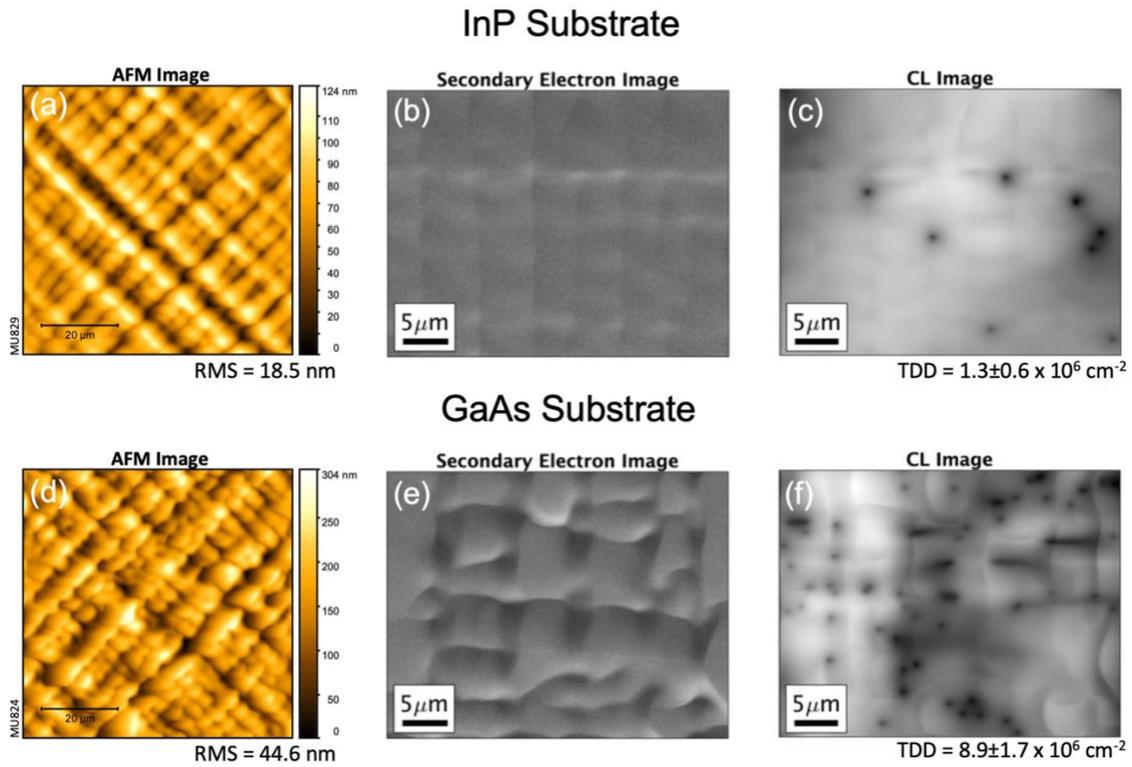

Figure 2. AFM, SEM, and CL images for graded buffers to the Ga$_{0.30}$In$_{0.70}$As lattice constant based on InP [(a)-(c)] and GaAs [(d)-(e)] substrates. The SEM and CL images are taken from the exact same place. The RMS roughness is listed below each AFM image. The threading dislocation density, averaged over a series of 8 images from each sample, is listed below each CL image.

## 2.2 Device Measurements

Next, we studied the behavior of the $Ga_{0.30}In_{0.70}As$ devices under the well-known one-sun (AM1.5G) reference spectrum and irradiance to benchmark their performance. Figure 3a shows the external and internal quantum efficiency (EQE and IQE) and Figure 3b shows light current density-voltage ($J$-$V$) curves, respectively, for these devices. The short wavelength cutoff in the QE results from absorption in the graded buffer. The bandgaps of these devices vary slightly, possibly due to reactor drift, variance in In/Ga incorporation efficiency on surfaces with varying roughness, or residual strain differences. Both devices show a high degree of carrier collection with peak IQE of ~95% for both devices. The open circuit voltage ($V_{OC}$) at one-sun is 45 mV larger in the InP-based device relative to the GaAs-based device, 0.200 vs. 0.155 V, which we attribute to the significantly lower TDD achieved on InP. The bandgap voltage offsets ($W_{OC} = E_G/q - V_{OC}$) at this irradiance are 0.38 and 0.44 V for InP and GaAs substrates, respectively. The fill factors (FF) for these cells are 61.5% and 54.5%, respectively.

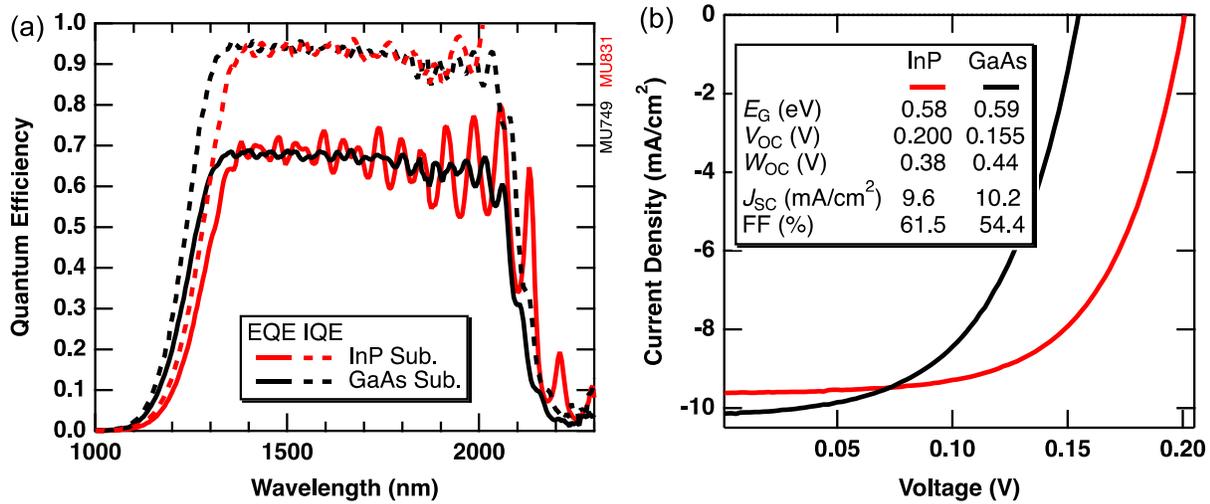

Figure 3. Quantum efficiency (a) and light J-V curves under a one-sun spectrum (b) for the devices based on InP and GaAs.

We then characterized these devices under high irradiance using a high-intensity pulsed solar simulator (HIPSS), to assess their performance under irradiance levels similar to thermophotovoltaics or laser power conversion applications. These measurements provide the dependence of $V_{OC}$ and FF on $J_{SC}$. As described below, we then use the EQE to map $J_{SC}$ and thus also $V_{OC}$ and FF onto the TPV spectrum of interest, effectively using the cell as its own reference cell to correct for the spectral mismatch between the spectrum of the simulator and the spectrum of interest. Fig. 4(a-c) shows select light $J$-$V$ curves, $V_{OC}$ and FF vs. $J_{SC}$, and dark $J$-$V$ curves, respectively, for both devices. The $V_{OC}$-$J_{SC}$ pairs from (b) are placed as open circles on the dark $J$-$V$ curves, to show the device behavior in absence of series resistance, which does not affect the $V_{OC}$ determination because no current is flowing. We fit the dark $J$-$V$ and HIPSS data to a double-diode optoelectronic device model[29] to derive more understanding from the data. The fits are shown as broken lines in panels (b) and (c), and the fit parameters are shown in Table I, where $J_{01}$ is the diode saturation current for the diode representing recombination in the quasi-neutral regions, $J_{0n}$ is the diode saturation current for the diode representing recombination in the space charge region, n is the diode ideality factor, and $R_S$ is the series resistivity. The $V_{OC}$ of the InP based device initially increases logarithmically with $J_{SC}$, as expected, although we note that the $V_{OC}$ begins to deviate from the model fit, likely due to device heating that occurs at higher irradiances. We measure a $V_{OC}$ of 0.386 V ($W_{OC}$ = 0.20 V) at 10 A/cm$^2$, under the highest irradiance tested. The model suggests that the $V_{OC}$ would be 0.401 V ($W_{OC}$ = 0.18 V) at this current density if not for the device heating. The FF peaks at 70.8% at 0.9 A/cm$^2$, and the product $V_{OC}$FF peaks at a current density of 3.0 A/cm$^2$, above which point the series resistance from the contacts causes significant reductions in fill factor.

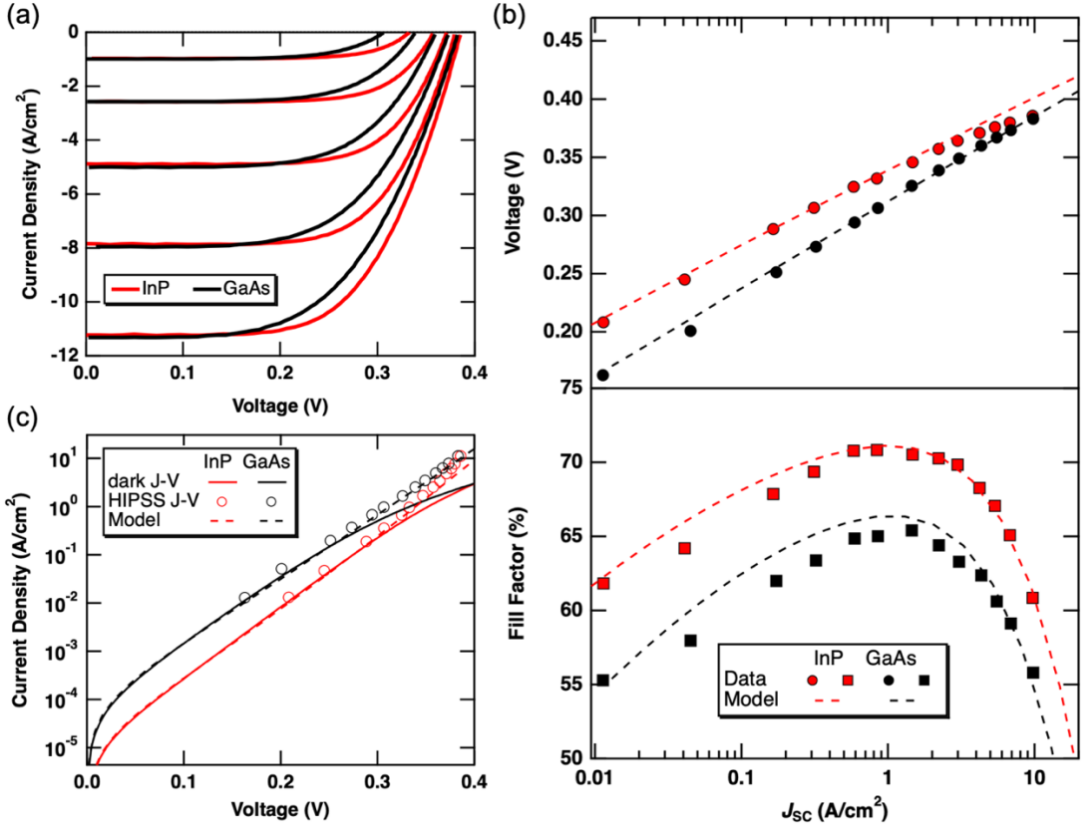

Figure 4. (a) High-intensity light $J$-$V$ curves measured on the <0.6 eV devices grown on InP and GaAs using a solar simulator with a pulsed Xe-arc lamp and GaAs filter. The light intensity was adjusted by varying the width of an aperture in front of the lamp. (b) $V_{OC}$ and fill factor as a function of $J_{SC}$ for these high-intensity measurements. (c) Dark J-V curves of the two devices. The $J_{SC}$-$V_{OC}$ pairs from (b) are plotted on the graph as open circles to show device behavior unaffected by series resistivity. Fits to a two-diode device model are plotted as broken lines in (b) and (c).

Table I – Fit parameters for the device model

|  | InP-based | GaAs-based |
| --- | --- | --- |
| $J_{01}$ (A cm$^{-2}$) | 8.4x10$^{-07}$ | 7.3x 10$^{-07}$ |
| $J_{0n}$ (A cm$^{-2}$) | 1.1x10$^{-05}$ | 7.5x10$^{-05}$ |
| $n$ | 1.3 | 1.3 |
| $R_S$ (mΩ cm$^2$) | 6.9 | 8.5 |

The performance of the GaAs-based device is understandably reduced compared to the InP based device because the $J_{0n}$ current is over 7x larger, but for that reason the $V_{OC}$ of the GaAs-based devices increases faster with current density relative to the InP based device. We estimate the ideality factor of the $J_{0n}$ diode in both devices to be 1.3, which is somewhat lower than commonly observed but within the expected range.[30] Devices utilizing compositionally graded buffers that are limited by recombination at dislocations at low current density often show an ideality factor of $1 < n < 2$. Notably, we have previously observed n = 1.5 in our metamorphic GaInAs junctions.[31] This reduced ideality factor means that the $V_{OC}$ vs. $J_{SC}$ curves do not converge in the current density range tested despite the fact that the GaAs based device has a lower $J_{01}$ current (in part because its bandgap is slightly larger). We measure a $V_{OC}$ of 0.383 V ($W_{OC}$ = 0.21 V) at 9.8 A/cm$^2$, and FF peaks at 65.4% at 1.5 A/cm$^2$. The $V_{OC}$FF product peaks at 4.3 A/cm$^2$ with $V_{OC}$ = 0.360 V and FF = 62.4%. $V_{OC}$FF peaks later for this device compared to the one on InP even though the series resistance was slightly lower. We attribute this fact to reduced heating in the GaAs based device, which is indicated by minimal deviation of the HIPSS $V_{OC}$ data from the best fit in Figure 4c. Thus the penalty that heating places on $V_{OC}$ and FF is reduced. We attribute the reduction in device temperature to the additional device thickness of the grade from GaAs to InP, which adds ~5 µm of thickness on top of the ~7 µm combined InP$_y$As$_{1-y}$ graded buffer and GaInAs device layer thickness of the InP-based device. This extra thickness presumably aids with heat dissipation in the device during pulsed measurement. We note that we had considered Auger recombination, which can affect low bandgap III-V devices,[32] as a cause of the $V_{OC}$ droop at high current density observed in the InP-based device, but the absence of droop at high current density in the GaAs-based device suggests that heating was indeed the cause.

**2.3 Thermophotovoltaic Efficiency Modeling**

Next, we used the high-intensity *J-V* measurements combined with measurements of the spectral reflectance of the devices to estimate TPV efficiency as a function of blackbody temperature. Figure 5a shows the normal reflectance as a function of wavelength on the left axis, with blackbody spectra at various temperatures overlaid on the right axis for reference. We did not measure the angular dependence of the reflectance, but note that near-normal measurements are generally a good predictor of results using analytical angle-integrated calculations.[33] Above bandgap, the reflectance is ~0.3 for both cells, typical of III-V cells without anti-reflection coating. The sub-bandgap reflectance is significantly higher in the InP-based device compared to the GaAs-based device. This difference implies that there is more sub-bandgap absorption in the GaAs-based device, which we suspect occurs in its thicker graded buffer, because the remainder of the structure is nominally identical. The absorption mechanism is possibly free carrier absorption and/or absorption caused by dislocations.[34-35] Given that the lattice mismatch between the substrate and device is much larger on GaAs (1.2% vs. 5%) the graded buffer on GaAs contains a considerably larger volume of dislocations (both threads and misfit dislocations) that may be responsible increased sub-bandgap absorption in this device.

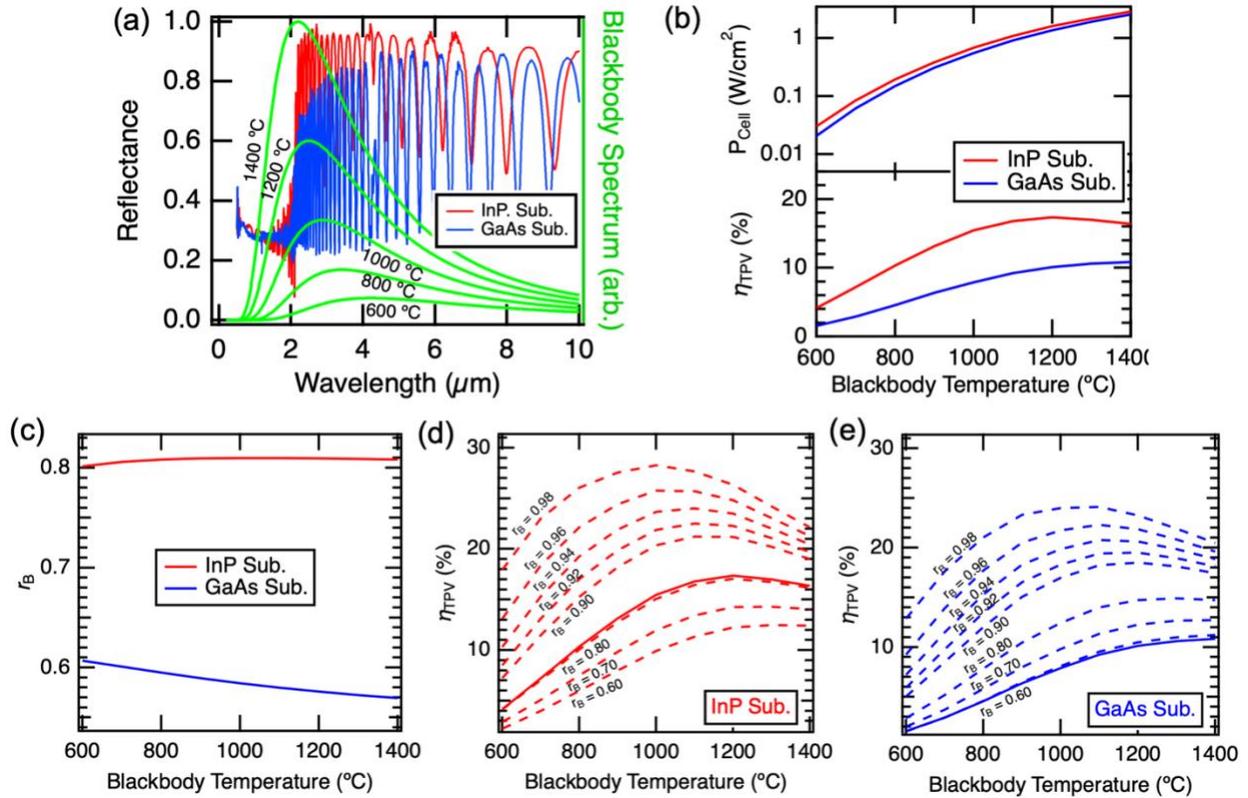

Figure 5. (a) Spectral reflectance of the <0.6 eV devices grown on InP and GaAs substrates. The blackbody spectra at various temperatures are plotted as green curves on the right axis. (b) Power produced by the cells (top) and $\eta_{TPV}$ (bottom) plotted as a function of blackbody temperature. (c) Power-weighted sub-bandgap reflectance for each cell as a function of blackbody temperature. (d) and (e) TPV efficiency vs. blackbody temperature at multiple values of average sub-bandgap reflectance ($r_B$) for the cell grown on the InP or the GaAs substrate.

Next we estimate $\eta_{TPV}$ as a function of blackbody temperature. The blackbody radiation spectrum is calculated using Planck's law. We use the equation:

$$\eta_{TPV} = \frac{P_{cell}}{P_I - P_R}, \quad (1)$$

where $P_{cell}=V_{OC}J_{SC}FF$ is the power generated by the cell, $P_I$ is the power incident on the cell at a given blackbody temperature assuming unity view factor, and $P_R$ is the power reflected by the cell back to the hypothetical emitter. $P_R$ is not counted as a loss because the energy is reabsorbed and then reradiated by the emitter.[17] $P_{cell}$ is estimated by first integrating the EQE of the cells with the blackbody spectrum to get $J_{SC}$, and then determining $V_{OC}$ and FF at that $J_{SC}$ by interpolating the data in Figure 4b. $P_R$ is determined by integrating the reflectance with the blackbody spectrum. Figure 5b plots $P_{cell}$ and $\eta_{TPV}$ as a function of blackbody temperature. The cell grown on InP provides more power at a given temperature because of its superior device metrics. As one point of comparison, the InP cell yields 1.09 W/cm² at 1100 °C, 18.4% more power than the 0.92 W/cm² provided by the GaAs cell. While this difference is reasonable given the increased TDD in the GaAs cell, $\eta_{TPV}$ in the InP cell is 83% higher relative to the GaAs efficiency (16.8% vs. 9.2%), due to the large penalty paid to the high sub-bandgap absorption in the GaAs cell. The peak efficiency for the InP cell is 17.4% at 1200 °C vs. 10.8% at 1400 °C for the GaAs cell. We note that application of an anti-reflection coating to these devices would increase these efficiencies by reducing the sub-bandgap absorption loss by ensuring more of the high energy photons are absorbed and converted on the first pass.

Improvement of the sub-bandgap reflectance is another promising route to improve the TPV efficiency of both devices. The power-weighted sub-bandgap reflectances ($r_B$) for the InP and GaAs-based cells hover around ~80% and ~60%, respectively, as shown in Figure 5c. We modeled the effect of $r_B$ on $\eta_{TPV}$ as a function of blackbody temperature for each cell as shown in Figures 5d and 5e. As expected, $\eta_{TPV}$ increases strongly with $r_B$ for both cells by reducing the value of the denominator in equation (1). We expect that removal of the graded buffer would significantly increase $r_B$ in these cells. Previously we demonstrated $r_B = 0.931$ in an inverted metamorphic tandem device at the $Ga_{0.70}In_{0.30}As$ lattice constant that had the graded buffer removed.[17] Integration of the reflectance from that cell with the 1000 °C blackbody temperature yields $r_B = 0.935$, providing a reasonable target for the present cells if a similar, removable graded buffer can be developed. If we achieve this value in an optimized device design that removes the grade, the efficiency of the InP and GaAs cells would peak at ~ 1100 °C at values of 24.0% and 20.7%, respectively. Devices using more sophisticated reflector designs such as an air gap have demonstrated $r_B$ up to 0.985,[18] implying even higher efficiencies are possible if those architectures could be incorporated in an optimized device.

**2.4 Laser Power Conversion Efficiency Modeling**

Another application for these devices is laser power conversion in the 2.0-2.3 µm atmospheric window. In this section, we estimate a laser power conversion efficiency assuming an EQE of 0.9 at 2 µm with a properly tuned anti-reflection coating (see IQE in Figure 3a). Assuming variable power source with a wavelength of 2.0 µm, we can calculate $J_{SC}$ from the following equation[36-37]:

$$J_{SC} = E_{tot} EQE(\lambda) \left(\frac{hc}{q\lambda}\right)^{-1} \quad (2)$$

where $E_{tot}$ is the total irradiance in W/cm$^2$ and $\frac{hc}{q\lambda} = 0.62$ eV. The laser power conversion efficiency is then calculated by:

$$\eta_{LPC} = \frac{P_{Cell}}{E_{tot}} \quad (3)$$

where $P_{Cell}$ is calculated as described as a function of $J_{SC}$ above for the TPV estimate. Figure 6 plots $\eta_{LPC}$ as a function of 2.0 µm irradiance.

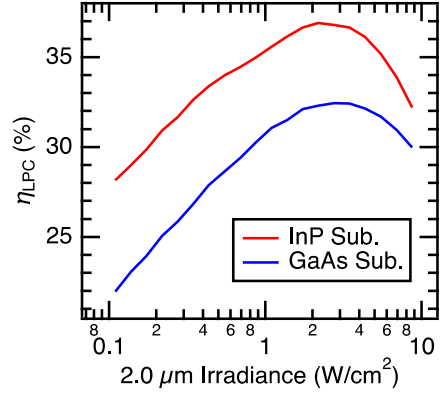

Figure 6. Estimated laser power conversion efficiency as a function of irradiance with a hypothetical 2.0 µm laser source.

The device grown on the InP substrate yields an estimated peak efficiency of 36.8% under an irradiance of 1.86 W/cm$^2$. The GaAs efficiency peaks at a higher current density presumably due to improved heat dissipation, yielding a 32.5% conversion efficiency at 2.81 W/cm$^2$. We note that use of multijunction devices could yield significant improvements by reducing series resistance losses.

## 3. Conclusion

We developed inverted metamorphic Ga$_{0.3}$In$_{0.7}$As photovoltaic converters with bandgaps below 0.60 eV grown on InP and GaAs substrates. These devices have application as thermophotovoltaics or converters for laser power beaming in the 2.0-2.3 µm atmospheric window. While the device grown on GaAs exhibits a higher threading dislocation density, the devices yield comparable $V_{OC}$s under high irradiance. The GaAs-based device has a lower sub-bandgap reflectance than the InP-based device, which we attribute to absorption in the thicker graded buffer in that device. We estimate these devices to yield 1.09 W/cm$^2$ vs. 0.92 W/cm$^2$, respectively, under a 1100 °C blackbody, with TPV efficiencies of 16.8 vs. 9.2%. Removal of the graded buffers from these devices should lead to significant increases in sub-bandgap reflectance and TPV efficiency. We estimate that the efficiencies would peak at ~ 1100 °C at values of 24.0% and 20.7% with reflectances previously achieved in other metamorphic devices with removed graded buffers. As LPCs, we estimated peak efficiencies of 36.8% and 32.5% for irradiation from an idealized 2.0 µm source.

## 4. Experimental Section

All devices were grown by organometallic vapor phase epitaxy (OMVPE) using standard precursors. The wafer curvature and surface reflectivity were monitored *in situ* using a multi-beam optical stress sensor. Substrates were either (100) InP:S or (100) GaAs:Si with a misorientation of 2° towards the (111)B plane. The device structures were shown in Fig. 1. The device on InP was grown with a InP:Si buffer, a 1 µm Ga$_{0.47}$In$_{0.53}$As:Si etch stop layer followed by a 0.5 µm InP:Si second etch stop layer, then a 0.2 µm Ga$_{0.47}$In$_{0.53}$As:Se front contact layer, where the element listed after the colon indicates the dopant. Next a In$_x$P$_{1-x}$As:Si step-graded buffer was grown with 0.5 µm steps and a strain grading rate of 0.33% strain/µm. The strain overshoot layer of the grade was grown 1.0 µm thick, and then the device layers were grown lattice-matched to the in-plane lattice constant of the overshoot. The total lattice mismatch between the device layers and the substrate is 1.2%. The front junction $Ga_{0.30}In_{0.70}As$ device layers and thicknesses are listed in Fig 1. The emitter doping density was nominally $1x10^{18}$ cm$^{-3}$ Se, while the Zn doping density in the base was measured by the capacitance-voltage method to be $2x10^{16}$ and $5x10^{16}$ cm$^{-3}$ in the devices grown on GaAs and InP, respectively, despite the molar fraction of dimethylzinc dopant being equal during each growth. It is likely that dopant incorporation rates are slightly different in each device due to differences in surface roughness. $InP_{0.64}As_{0.36}$ doped $\sim 1x10^{18}$ cm$^{-3}$ with Si or Zn was used to passivate the front and rear surfaces of the device, respectively. The device grown on GaAs substrate used a $Ga_{0.5}In_{0.5}P$:Se etch stop and GaInNAs:Se [38] front contact layer, followed by a $Ga_xIn_{1-x}P$:Si graded buffer from GaAs to InP, and then the same $In_xP_{1-x}As$:Si graded buffer used in the cell grown on InP. The $Ga_xIn_{1-x}P$ buffer was graded in 0.25 µm steps at 1% strain/µm from $x_{Ga}$ = 0.50-0.20, and then 0.5% strain/µm from $x_{Ga}$ = 0.20-0.00 (InP) using 0.125 µm steps.[39] The total lattice mismatch from substrate to device is 5.0%. The dislocation density of the graded buffers was analyzed by CL in a scanning electron microscope. Panchromatic CL intensity images were collected using a parabolic mirror to direct the emission due to radiative transitions to a liquid nitrogen cooled Ge photodiode. An accelerating voltage of 15kV and electron beam current of $\sim$ 0.7 nA were used to generate excess carriers in the samples. We grew separate structures for CL analysis that stopped at the $InP_{0.64}As_{0.36}$ layer, which had a bandgap ($E_G$) of 0.93 eV, because the cooled Ge CL detector ($E_G$ = 0.66 at room temperature) cannot detect emission from $Ga_{0.30}In_{0.70}As$. We assume that the TDD measured in the final grade layer is roughly the same as that in the $Ga_{0.30}In_{0.70}As$ device layers, which are grown lattice-matched to the in-plane lattice constant of the graded buffer.

After growth, devices were made by first electroplating the Au rear contact, which also serves as the rear reflector, bonding that contact to a Si substrate handle with a low viscosity epoxy, and then removing the substrate using wet chemical etchants that do not react with the etch stop layers. Then the front contact concentrator grid was defined by photolithography and electroplated onto the front contact layer. Square 0.116 cm$^2$ devices were defined by photolithography and mesas were isolated using selective wet etching.[40] We measured the wavelength-dependent EQE on a calibrated system with a 2300 nm GaInAs reference detector. IQE was calculated from EQE and the measured wavelength-dependent reflectance, $r(\lambda)$ using the equation $IQE(\lambda) = EQE(\lambda)/[1-r(\lambda)]$. We measured *J-V* in the dark and in the light at 25 °C under an XT-10 system with a Xe-arc lamp. Due to the lack of a calibrated reference cell with the same bandgap as the cell, we used the cell as its own reference, adjusting the distance between the lamp and the cell until the measured $J_{SC}$ matched that calculated by integration of the EQE with the ASTM G173 AM1.5 Global spectrum. Based on comparisons between EQE and *J-V* measurements for higher bandgap single-junction cells for which we do have good reference cells, we estimate the systematic uncertainty to be <3%. We note that AM1.5G is not an appropriate reference spectrum for TPV and LPC applications, but it is useful as a well-defined reference. We characterized the devices under high intensity irradiation using a high-intensity pulsed solar simulator (HIPSS) with a Xe arc lamp and a GaAs filter to eliminate radiation below 880 nm. The approximate spectrum is shown in ref. [17]. The irradiance was controlled by varying the size of an aperture in front of the lamp.

The reflectance was measured in a normal configuration using a Bruker V70 Fourier-transform infrared (FTIR) spectrometer and a Hyperion 2000 microscope. An aperture of spot size approximately 0.01 mm$^2$ was used and the spot was placed completely between the front contract gridlines of the device. The three spectral ranges come from three different combinations of light sources/beam splitters/detectors with the visible range coming from tungsten lamp/Quartz beam splitter/Silicon detector, the Near IR from Tungsten lamp/Quartz beam splitter/HgCdTe detector and the Mid IR coming from Globar/KBr beam splitter/ HgCdTe detector. All measurements were taken at room temperature and referenced to a NIST standard that has known reflectance to within +/-0.4% absolute.

## 5. Acknowledgments

The authors would like to thank Michelle Young for device growth and Sarah Collins for device processing. This work was authored by the National Renewable Energy Laboratory (NREL), operated by Alliance for Sustainable Energy, LLC, for the U.S. Department of Energy (DOE) under Contract No. DE-AC36-08GO28308. This work was supported by the Laboratory Directed Research and Development (LDRD) Program at NREL. Support for E.J.T. and T.D. was provided by the University of Wisconsin-Madison, Office of the Vice Chancellor for Research and Graduate Education with funding from the Wisconsin Alumni Research Foundation. The views expressed in the article do not necessarily represent the views of the DOE or the U.S. Government. The U.S. Government retains and the publisher, by accepting the article for publication, acknowledges that the U.S. Government retains a nonexclusive, paid-up, irrevocable, worldwide license to publish or reproduce the published form of this work, or allow others to do so, for U.S. Government purposes.